\documentclass[prd,preprint,showpacs,english,nofootinbib,floatfix]{revtex4}
\usepackage[T1]{fontenc}
\usepackage[latin1]{inputenc}
\usepackage{babel}
\usepackage{graphicx}
\usepackage{epsf}
\usepackage{setspace}

\begin{document}
\def\beqn{\begin{eqnarray}}
\def\eeqn{\end{eqnarray}}
\def\nn{\nonumber}
\preprint{WM-05-115}
%
%
\title{Photoproduction of the $\Theta^+$ resonance on the nucleon
in a Regge model}
\author{H. Kwee}
\affiliation{Department of Physics, College of William and Mary,
Williamsburg, VA 23187, USA}
\author{M. Guidal}
\affiliation{Institut de Physique Nucl{\'e}aire Orsay, F-91406 Orsay, France}
\author{M.V. Polyakov}
\affiliation{Institut f\"ur Theoretische Physik II, Ruhr-Universit\"at Bochum,
D-44780 Bochum, Germany}
\affiliation{Petersburg Nuclear Physics Institute, 188350, Gatchina, Russia}
\author{M. Vanderhaeghen}
\affiliation{Jefferson Laboratory, Newport News, VA 23606, USA}
\affiliation{Department of Physics, College of William and Mary,
Williamsburg, VA 23187, USA}
\date{\today}
\begin{abstract}
We estimate the reaction mechanisms for the photoproduction of the
$\Theta^+(1540)$ resonance on the nucleon, through
$K$ and $K^*$ Regge exchanges.
We compare the size of the cross sections for the
$\gamma n \to K^- \Theta^+$ and $\gamma p \to \bar K^0 \Theta^+$ reactions, and
investigate their sensitivity to the spin-parity assignments $J^P =
\frac{1}{2}^\pm, \frac{3}{2}^\pm$ for the $\Theta^+$ resonance.
The model allows to estimate the cross sections corresponding with a given
upper bound on the width of the $\Theta^+$.
Within this model, the cross sections on
the neutron are found to be around a factor 5 larger than the ones
on the proton, due to the presence of charged $K$ exchange for the reaction on
a neutron target. Furthermore, the photon asymmetry is found to display a
pronounced sensitivity to the parity of the $\Theta^+$, making it a
very promising observable to help determining the quantum numbers of
the $\Theta^+$ resonance.
\end{abstract}
\maketitle

\section{Introduction}

Since the announcement of the first claim of experimental evidence for the
$S = 1$ baryon resonance by Spring 8~\cite{spring-8} which is well known
as the $\Theta^+$ today, there has been a great number of experiments
reporting the observation of the
$\Theta^+$~\cite{diana,clas,saphir,neutrino,clas-2,hermes,zeus,cosy,svd}.
These experiments represent a great variety of nuclear and particle
physics experiments.  The first experiment~\cite{spring-8} is a
photoproduction process on carbon nuclei.  The subsequent experiments
searched for the $\Theta^+$ in photoproduction on the
nucleon~\cite{clas,saphir,clas-2}, electroproduction on the
nucleon~\cite{hermes,zeus}, and using kaon beams~\cite{diana}, neutrino
beams~\cite{neutrino} and hadron beams~\cite{cosy,svd}.  All of these
experiments have some weaknesses which made any single claim for the
$\Theta^+$ not robust. If however taken together these experiments present
significant claim for the $\Theta^+$.
Most of the criticism arises due to the low statistics of the data of
most experiments, uncertainty in background estimates, specific
angle cuts applied for the data analysis~\cite{diana,saphir,clas-2}, mixed
strangeness~\cite{neutrino,hermes,svd} and certain nuclear effect
corrections~\cite{spring-8,clas}.  In addition there is also a claim of
observation for $S = -2$ exotic pentaquarks~\cite{na49}, even though this
claim cannot be reproduced~\cite{fischer, WA89}.

Criticism for the $\Theta^+$ claim also arises because a fairly large
number of experiments~\cite{BES,BaBar,Belle,ALEPH,HERA-B,SPHINX,HyperCP,CDF,FOCUS,Belle2,PHENIX} did not find any evidence for the $\Theta^+$.
Most of these experiments are high energy-high statistics experiments and
they can fall into two categories.  The first category are the $e^+ e^-$
experiments~\cite{BES,BaBar,Belle,ALEPH} and the second category are the
hadron beam experiments~\cite{HERA-B,SPHINX,HyperCP,CDF}.  The $e^+ e^-$
experiments do not present credible challenge to the existence of the
$\Theta^+$ since it is very unlikely to produce the $\Theta^+$ from these
experiments.  The hadron beam experiments present more serious rebuttal
evidence to worry about, but like the $e^+ e^-$ experiments, without
obvious production mechanism, the rebuttal is not very convincing.  It is
also important to point out that the SAPHIR result~\cite{saphir} has been
ruled out by the recent CLAS g11 experiment~\cite{clas-3}.  Details of
analysis for the positive results as well as the negative results for
the pentaquark experiments can be found in~\cite{hicks}.

From the theoretical point of view, QCD does not prohibit the existence
of pentaquark states.  If there are any prejudices, we should expect
the existence of the exotic states such as dibaryons, dimesons, glueballs
and pentaquarks.  For the pentaquark, the problems are where to look for the
states and how to distinguish them from the ordinary baryon resonances.
Also if the states are very wide, the search might be futile since we will
not be able to distinguish the resonances from the background.  After
years of searches with no results, the community seemed to give up when
PDG after 1986~\cite{pdg} dropped the section on searches for pentaquarks.
Recent interest for pentaquarks was sparked again after the announcement
from Spring 8 experiment~\cite{spring-8}.  The experiments were conducted
with the guidance of the chiral quark soliton model estimate by Diakonov
{\it et. al.}~\cite{DPP97}, which predicted the existence of a narrow
baryon resonance with strangeness ($S = 1$) at mass around 1530 MeV. 
Critics for this paper can be found in~\cite{Wei98,Cohen03}.  The paper
also predicted that the resonance has spin 1/2 and positive parity as
predicted by subsequent chiral soliton model
works~\cite{Hos03,Wall03,Bor03,Pras03}.  A naive constituent quark model
will give a negative parity state~\cite{us1}, even though it is also
possible to have a positive parity state in a quark model if one
introduces a P-wave in the spatial wavefunction~\cite{JW03,us2,Sta03}.  As
for the spin of the $\Theta^+$, all theory paper to our knowledge predicts
1/2 and consider spin-3/2 to be a heavier excitation state, see {\it e.g.}
Ref.~\cite{ClDudek03}.

The most puzzling issues from the theoretical point of view right now is
the narrow width of the $\Theta^+$.  The chiral quark soliton model work
by Diakonov {\it et. al}~\cite{DPP97} did naturally predict an anomalously
narrow width for the $\Theta^+$.  Some attempts also have been made in
constituent quark models to explain the narrow width, {\it
e.g.}~\cite{us3,Sta03,Caps03}.  For an early review on these theoretical
issues see~\cite{Jen03}.

Clearly, to address the issues mentioned above, {\it i.e.} existence,
spin-parity assignment and width of the pentaquark, more dedicated experiments
are needed.  Production mechanisms are a key aspect
in this study since it has been argued that certain
processes will not be effective to produce the pentaquark.  Many efforts
have been made in the past to study the $\theta^+$ photo- and electroproduction
processes~\cite{Liu03,Nam03,Oh04,Zhao04,NaTs04,ZhAK04,ClZh04,NaLo04,TMart,Roberts,pire},
even though most of the cross section predictions have been ruled out by
the recent JLAB g10 and g11 experiments~\cite{clas-3,clas-4}.

In this paper we make another effort to study the photoproduction
mechanism for the $\Theta^+$ using a Regge model,
which has been found to successfully describe the main features of
$K \Lambda$ and $K \Sigma$ photoproduction on the nucleon at {\it c.m.}
energies above 2 GeV~\cite{GLV97,GLV00}. In section II we discuss
the theory of Regge exchange mechanism.  We also discuss the relation of
the width to the photoproduction cross section of the $\Theta^+$.
In view of upcoming high resolution experiments~\cite{halla},  such a
link is needed to translate quantitatively an upper limit on
the photoproduction cross section into an upper bound on the
$\Theta^+$ width.
In section III we present the results of our calculation for the $\Theta^+$
photoproduction cross sections.  We also
study photon asymmetry and decay angular distribution of the $\Theta^+$
photoproduction.  We close with our conclusion in section IV.

\section{Regge model}

\begin{figure}[ht]
\epsfxsize=15cm
\centerline{\epsffile{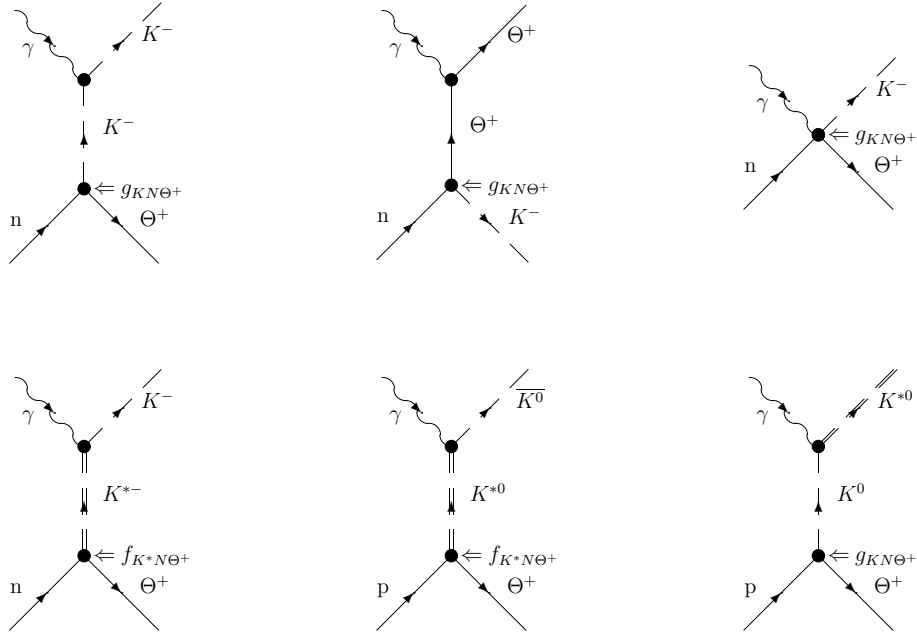}}
\vspace{-7.5cm}
\caption{The different Regge exchange contributions considered to describe
the photoproduction reactions with $\Theta^+$ final state on a nucleon.}
\label{fig:diag}
\end{figure}

At sufficiently high energies, above the nucleon resonance region,
strangeness photo- and electroproduction reactions (e.g. $\gamma p \to K^+
\Lambda, K^+ \Sigma$) at forward angles are dominated by $K$ and $K^*$
Regge exchanges as has been proposed long time ago in Ref.~\cite{LMR73},
and studied in detail in view of numerous recent strangeness photo- and
electroproduction data in Refs.~\cite{GLV97,GLV00}.  It was found in those
works that the simple Regge model for open strangeness electromagnetic
production reactions in terms of $K$ and $K^*$ exchanges provides an
economical description and simple explanation of the forward angle data
for total {\it c.m.} energy $W > 2$~GeV.  It surprisingly reproduces the
gross features of the data, even for $W < 2$~GeV, hinting that a sort of
reggeon-resonance duality is at work.
\newline
\indent
In this work, our aim is to extend this model to the description of the
process
\begin{eqnarray}
\gamma (q) \,+\, N (p_N) \, \longrightarrow \, K (p_K) \,+\, \Theta^+
(p_\Theta) \, .
\label{eq:reaction}
\end{eqnarray}
The Mandelstam variables for this process are given by $s \equiv (p_N +
q)^2$, $t \equiv (q - p_K)^2$, and $u \equiv (q - p_\Theta)^2$, satisfying
$s + t + u = M_N^2 + m_K^2 + M_\Theta^2$, with $M_N$ the nucleon mass,
$m_K$ the kaon mass, and $M_\Theta$ the mass of the $\Theta^+$.
In this work we take as value for the mass of the $\Theta^+$ the value
$M_\Theta = 1.54$~GeV, consistent with the experiments of
Refs.~\cite{spring-8,diana,clas,saphir,clas-2,hermes,neutrino,svd,cosy}.
Our description of the reaction (\ref{eq:reaction}) in terms of
reggeized $K$ and $K^*$ $t$-channel exchanges is aimed at the region
of large $s$ ($W = \sqrt{s} > 2$~GeV) and small $-t$ ($-t << s$).
We discuss subsequently the $t$-channel $K$ and $K^*$ Regge exchange
processes for reaction~(\ref{eq:reaction}) as is shown in
Fig.~\ref{fig:diag}.

\subsection{$K$ Regge exchange}

\subsubsection{Spin-1/2 $\Theta^+$ Amplitude}

The construction of the reggeized amplitude for the charged $K$-exchange
process to the $\gamma n \to K^- \Theta^+$ reaction, as shown in
Fig.~\ref{fig:diag}, amounts to replace the usual Feynman pole propagator
for the kaon by a `Regge propagator' function, depending on both $s$ and
$t$, i.e. $\mathcal{P}^{K}_{Regge}(s,t)$ as follows~:
\begin{eqnarray}
\frac{1}{t-m_{K}^2}\Longrightarrow
\mathcal{P}^{K}_{Regge}(s,t) \,=\,
\left({\frac{s}{s_0}}\right)^{\alpha_{K}(t)}
\frac{\pi \alpha_{K}'}{\sin(\pi \alpha_{K}(t))}\;
\frac{\mathcal{S}_K+e^{-i\pi \alpha_{K}(t)}}{2}\;
\frac{1}{\Gamma(1+\alpha_{K}(t))}\, .
\label{eq:kaonregge}
\end{eqnarray}
Such a Regge propagator function effectively takes into account the
exchange of high-spin particles (in the $t$-channel) which lie on the $K$
Regge trajectory $\alpha_K(t) = \alpha_K^0 + \alpha'_K \cdot t $.  For the
$K$, we use a standard linear trajectory in Eq.~(\ref{eq:kaonregge})~:
\begin{eqnarray}
\alpha_K(t)= 0.7 \,(t-m^2_K) \;,
\label{eq:kaontraj}
\end{eqnarray}
which was previously used in the analysis of strangeness photoproduction
reactions at high energies~\cite{GLV97}.  Furthermore, in
Eq.~(\ref{eq:kaonregge}), the mass scale $s_0$ is typically taken as $s_0
= 1\;\mathrm{GeV}^2$, and $\mathcal{S} =\pm 1$ is the signature of the
trajectory~\cite{Col77}.  For the kaon trajectory, the states with
$J^P = 0^-, 2^-, 4^-, ...$ correspond with $\mathcal{S} = +1$,
whereas the states with
$J^P = 1^+, 3^+, 5^+, ...$ correspond with $\mathcal{S} = -1$.
The gamma function
$\Gamma(1 + \alpha(t))$ suppresses poles of the propagator in the
unphysical region.  As is well known from Regge theory~\cite{Col77},
trajectories can be either non-degenerate or degenerate.  A degenerate
trajectory is obtained by adding or subtracting the two non-degenerate
trajectories with the two opposite signatures.  As can be seen from the
numerator of Eq.~(\ref{eq:kaonregge}), this leads to trajectories with
either a rotating ($e^{-i\pi\alpha(t)}$) or a constant (1) phase.
In line with the finding of Ref.~\cite{GLV97} for the charged $K$
trajectory entering in strangeness photoproduction processes, we use the
rotating phase in the following.  This corresponds with the so-called
strong degeneracy assumption \cite{Col77} in Regge terminology, and
amounts to replace in Eq.~(\ref{eq:kaonregge}) the factor $
(\mathcal{S}_K+e^{-i\pi \alpha_{K}(t)}) / 2$ by $e^{-i\pi \alpha_{K}(t)}$.
One can easily verify that the Regge propagator reduces to the Feynman
propagator $1/(t-m_K^2)$ if one approaches the first pole on a trajectory
(i.e. when taking $t \rightarrow m_K^2$ in Eq.~(\ref{eq:kaonregge})).
\newline
\indent
In order to calculate the $K$ Regge exchange contribution to the
$\Theta^+$ photoproduction amplitude as shown in Fig.~\ref{fig:diag}, we
have to specify the $K N \Theta^+$ vertex function.  We subsequently
study this vertex for the cases of spin and parity
assignments $1/2^\pm$ and $3/2^\pm$ of the $\Theta^+$ resonance.
\newline
\indent
\begin{itemize}
\item
$J^P = \frac{1}{2}^+$ \\
For the spin parity assignment of the $\Theta^+$ resonance given by $J^P
= \frac{1}{2}^+$, the $K N \Theta^+$ vertex can be written as~:
\begin{eqnarray}
\mathcal{L}_{K N \Theta} \,=\,
i \, g_{K N \Theta} \, \left( K^\dagger \bar \Theta \gamma_5 N
\,+\,  \bar N \gamma_5 \Theta K \right)\, ,
\label{eq:lagrpospar}
\end{eqnarray}
where $K$ and $N$ are the kaon and nucleon isospin doublet fields
respectively, and where $\Theta$ is the $\Theta^+$ isosinglet field.  The
Lagrangian of Eq.~(\ref{eq:lagrpospar}) corresponds with the $\Theta^+$
being a $p$-wave resonance in the $K N$ system.  With this Lagrangian, the
decay width $\Gamma_{\Theta \to K N}$ is given by~:
\begin{eqnarray}
\Gamma_{\Theta \to K N} \,=\, \frac{g_{K N \Theta}^2}{2 \pi} \,
\frac{|\bar p_K|}{M_{\Theta}} \left(\sqrt{\bar p_K^2 + M_N^2} -
M_N\right),
\label{eq:widthpos}
\end{eqnarray}
where $|\bar p_K| \simeq 0.267$~GeV is the kaon momentum in the rest frame
of the $\Theta^+$.  To extract a value for $g_{K N \Theta}$, we need the
experimental information of the width $\Gamma_{\Theta \to K N}$, which is
not known precisely at this moment but whose measurement is the subject of
several planned dedicated experiments, e.g. Refs.\cite{halla}.  To provide
numerical estimates in this work, we will use $\Gamma_{\Theta \to K N}$ =
1 MeV as the value for the width, which is consistent with the upper
limit for the width derived from elastic $KN$ scattering~\cite{Cahn:2003wq}.
Evaluating Eq.~(\ref{eq:widthpos}) with $\Gamma_{\Theta \to K N}$ = 1 MeV,
we then extract the value $g_{K N \Theta} \simeq 1.056$, which will be
used in all of the following estimates for $J^P = \frac{1}{2}^+$ $\Theta^+$.
\newline
\indent
\item
$J^P = \frac{1}{2}^-$ \\
For the spin parity assignment of the $\Theta^+$ resonance given by $J^P
= \frac{1}{2}^-$, the $K N \Theta^+$ vertex can be written as~:
\begin{eqnarray}
\mathcal{L}_{K N \Theta} \,=\, g_{K N \Theta} \,
\left( K^\dagger \bar \Theta N \,+\,  \bar N \Theta K \right)\, ,
\label{eq:widthneg}
\end{eqnarray}
which corresponds with the $\Theta^+$ being a $s$-wave resonance in the $K
N$ system.  In this case, the decay width $\Gamma_{\Theta \to K N}$ is
given by~:
\begin{eqnarray}
\Gamma_{\Theta \to K N} \,=\, \frac{g_{K N \Theta}^2}{2 \pi} \,
\frac{|\bar p_K|}{M_{\Theta}} \left(\sqrt{\bar p_K^2 + M_N^2} +
M_N\right).
\end{eqnarray}
Using $\Gamma_{\Theta \to K N}$ = 1 MeV as the value for the width, we
obtain $g_{K N \Theta} \simeq 0.1406$, which will be used in the
following estimates for $J^P = \frac{1}{2}^-$ $\Theta^+$.
\end{itemize}
\indent
Having specified the Regge propagator and $K N \Theta$ vertex function, we
can construct the gauge invariant reggeized charged $K$-exchange amplitude
for the $\gamma n \to K^- \Theta^+$ process for the spin-1/2 $\Theta^+$ as~:
\begin{eqnarray}
{\mathcal M}_{K} \left(\gamma n \to K^{-} \Theta^+ :
J^P_\Theta = \frac{1}{2}^+ \right) &=&\,
e \, g_{K N \Theta} \, \cdot \, \mathcal{P}^{K}_{Regge}(s, t) \cdot
\varepsilon_\mu (q, \lambda) \, \nonumber \\
&\times&
\left[{F_K}(t) \,\cdot\, (2 p_K - q)^{\mu} \,\cdot\, \bar{\Theta} \,
\gamma^{5} \, N \right. \nonumber \\
&\;\;\;-& \; \left. {F_\Theta}(u) \,\cdot\, (t-m^2_K) \, \cdot \bar{\Theta}
\, \gamma^\mu \frac{(\gamma \cdot p_u + M_\Theta)}{u - M^2_\Theta} \gamma^5
\, N   \right. \nonumber \\
&\;\;\;+& \; \left. 2 p_K^\mu \,\cdot\, ({\hat F}(s,t,u) - F_K(t))
\,\cdot\, \bar{\Theta} \, \gamma^{5} \, N\right. \nonumber \\
&\;\;\;-& \; \left.  \left(\frac{t - m_K^2}{u - m_\Theta^2}\right)
\,\cdot\, 2 p_\Theta^\mu \,\cdot\, \{{\hat F}(s,t,u) - F_\Theta(u)\}
\cdot\, \bar{\Theta} \, \gamma^{5} \, N \right] \, , \nonumber \\
\label{eq:amplkaon_gankm_pospar} \\
{\mathcal M}_{K} \left(\gamma n \to K^{-} \Theta^+ :
J^P_\Theta = \frac{1}{2}^- \right) &=&\,
( - i) \, e \, g_{K N \Theta} \, \cdot \, \mathcal{P}^{K}_{Regge}(s, t) \cdot
\varepsilon_\mu (q, \lambda) \, \nonumber \\
&\times&
\left[F_K(t) \,\cdot\, (2 p_K - q)^{\mu} \,\cdot \, \bar{\Theta} \, N
\right. \nonumber \\
&\;\;\;-& \; \left. {F_\Theta}(u) \,\cdot\, (t-m^2_K) \, \cdot
\bar{\Theta} \, \gamma^\mu \frac{(\gamma \cdot p_u + M_\Theta)}{u -
M^2_\Theta} \, N \right. \nonumber \\
&\;\;\;+& \; \left. 2 p_K^\mu \,\cdot\, ({\hat F}(s,t,u) - F_K(t))
\,\cdot\, \bar{\Theta} \, N \right. \nonumber \\
&\;\;\;-& \; \left.  \left(\frac{t - m_K^2}{u - m_\Theta^2}\right)
\,\cdot\, 2 p_\Theta^\mu \,\cdot\, \{{\hat F}(s,t,u) - F_\Theta(u)\}
\,\cdot\, \bar{\Theta} \, N \right] \, , \nonumber \\
\label{eq:amplkaon_gankm_negpar}
\end{eqnarray}
where $\varepsilon_\mu(q, \lambda)$ is the photon polarization vector with
photon polarization $\lambda = \pm 1$.
To evaluate the Regge vertex functions (Regge residues) away from the pole
position, we include
the form factors $F_K(t)$, $F_\Theta(u)$, and $\hat F$
in the amplitude formulas above.
In our calculations, we choose monopole forms for $F_K$ and $F_u$~:
\begin{eqnarray}
F_K(t) &\,=\,& \left(1 + \frac{-t + m_K^2}{\Lambda^2}\right)^{-1} \, , \\
F_\Theta(u) &\,=\,& \left(1 + \frac{-u + M_\Theta^2}
{\Lambda^2} \right)^{-1} .
\end{eqnarray}
For the cut-off $\Lambda$, we choose a typical hadronic scale of
$\Lambda = 1$ GeV.
In Eqs.~(\ref{eq:amplkaon_gankm_pospar},\ref{eq:amplkaon_gankm_negpar}),
the terms proportional to $F_K$ are the $t$-channel process (top left
diagram in  Fig.~\ref{fig:diag}), whereas
the terms proportional to $F_u$
originate from the process with the $\Theta^+$ in the $u$-channel
(top middle diagram in Fig.~\ref{fig:diag}).
In the Regge approach, this gauge restoring term is reggeized in the same
way as for the $t$-channel process, see e.g. Refs.~\cite{Jon81,GLV97}.  In
particular, one notices that at the $K$ pole, the pre-factors $(t -
m_K^2)$ in the second terms of
Eqs.~(\ref{eq:amplkaon_gankm_pospar},\ref{eq:amplkaon_gankm_negpar})
exactly compensate the Regge function $\mathcal{P}^{K}_{Regge}$, reducing
these contributions to standard $u$-channel pole terms.  For the $\gamma p
\to K^+ (\Lambda, \Sigma)$ reactions, it has been shown in
Ref.~\cite{GLV00} that this gauge invariant reggeization procedure (for
the case where the form factors are absent), by
restoring the gauge invariance of the $t$-channel charged kaon exchange
process through proper reggeization of the $s$-channel (for $K^+$) or
$u$-channel (for $K^-$) processes, is a key to reproduce several
strangeness photo- and electroproduction observables.  The third and
fourth terms which contain $\hat F$ are
contact terms which are required by gauge invariance when
including form factors. The only restriction we applied
to these terms is they should not have poles at $t = m_K^2$ and $u =
m_\Theta^2$.  With this restriction, we can choose the form~:
\begin{eqnarray}
\hat F(s,t,u) =
F_K(t) + F_\Theta(u) - F_K(t)\,\cdot\,F_\Theta(u).
\end{eqnarray}
Note that $\hat F - F_K(t)$ is proportional to $(t - m_K^2)$,
whereas $\hat F - F_\Theta(u)$ is proportional to
$(u - M_\Theta^2)$, thus canceling the poles in the contact terms.

\subsubsection{Spin-3/2 $\Theta^+$ Amplitude}

\begin{itemize}
\item $J^P = \frac{3}{2}^+$ \\
For the spin parity assignment of the $\Theta^+$ resonance given by $J^P
= \frac{3}{2}^+$, the $K N \Theta^+$ vertex can be written as~:
\begin{eqnarray}
\mathcal{L}_{K N \Theta} \,&=&\,
 \frac{g_{K N \Theta}}{m_K} \, \left\{ \bar \Theta^\alpha g_{\alpha \beta}
N \left(\partial^\beta K \right)
\,+\,  \bar N \Theta^\alpha g_{\alpha \beta} \left(\partial^\beta
K^\dagger\right) \right\}\, . \nonumber
\label{eq:lagr3pospar}
\end{eqnarray}
The Lagrangian of Eq.~(\ref{eq:lagr3pospar}) corresponds with the
$\Theta^+$ being a $p$-wave resonance in the $K N$ system.
In this case, the decay width $\Gamma_{\Theta \to K N}$ is given by~:
\begin{eqnarray}
\Gamma_{\Theta \to K N} \,&=&\, \frac{g_{K N \Theta}^2}{2 \pi} \,
\frac{|\bar p_K|}{M_{\Theta}} \, \frac{|\bar p_K|^2}{3m_K^2}
\left(\sqrt{\bar p_K^2 + M_N^2} + M_N\right). \nonumber
\label{eq:width3pos}
\end{eqnarray}

Again using $\Gamma_{\Theta \to K N}$ = 1 MeV as value for the width, we
obtain $g_{K N \Theta} \simeq 0.4741$, which will be used in the
following estimates for $J^P = \frac{3}{2}^+$ $\Theta^+$.

\item $J^P = \frac{3}{2}^-$ \\
For the spin parity assignment of the $\Theta^+$ resonance given by $J^P
= \frac{3}{2}^-$, the $K N \Theta^+$ vertex can be written as~:
\begin{eqnarray}
\mathcal{L}_{K N \Theta} \,&=&\, \frac{g_{K N \Theta}}{m_K} \,
\left\{ \bar \Theta^\alpha \gamma_5 N g_{\alpha \beta}
\left(\partial^\beta K \right)
\,+\,  \bar N \gamma_5 \Theta^\alpha g_{\alpha \beta} \left(\partial^\beta
K^\dagger\right) \right\}\, . \nonumber
\label{eq:lagr3negpar}
\end{eqnarray}
The Lagrangian of Eq.~(\ref{eq:lagr3negpar}) corresponds with the
$\Theta^+$ being a $d$-wave resonance in the $K N$ system.
In this case, the decay width $\Gamma_{\Theta \to K N}$ is given by~:
\begin{eqnarray}
\Gamma_{\Theta \to K N} \,&=&\, \frac{g_{K N \Theta}^2}{2 \pi} \,
\frac{|\bar p_K|}{M_{\Theta}} \, \frac{|\bar p_K|^2}{3m_K^2} \left(\sqrt{\bar
p_K^2 + M_N^2} - M_N\right) . \nonumber
\label{eq:width3neg}
\end{eqnarray}

Finally, using $\Gamma_{\Theta \to K N}$ = 1 MeV as value for the width, we
obtain $g_{K N \Theta} \simeq 3.558$, which will be used in the
following estimates for $J^P = \frac{3}{2}^-$ $\Theta^+$.
\end{itemize}

Having specified the $K N \Theta$ vertex function for spin-3/2, we
can construct the gauge invariant reggeized charged $K$-exchange amplitude
for the $\gamma n \to K^- \Theta^+$ process for the $3/2^-$ $\Theta^+$ as~:

\begin{eqnarray}
{\mathcal M}_{K} \left(\gamma n \to K^{-} \Theta^+ : J^P_\Theta =
\frac{3}{2}^+\right)  &=&\,
\frac{i \, e \, g_{K N \Theta}}{m_K} \, \cdot \,
\mathcal{P}^{K}_{Regge}(s, t) \cdot \varepsilon_\mu (q, \lambda) \nonumber
\\
&\times&
\left[F_K(t) \,\cdot\, (2 p_K - q)^{\mu} \,\cdot\, (p_K - q)^{\alpha}
\,\cdot\, \bar{\Theta}_{\alpha} \, N \right. \nonumber \\
&\;\;\;-& \; \left. F_\Theta(u) \,\cdot\, (t-m^2_K) \, \cdot
\bar{\Theta}_{\alpha} \, \gamma^{\alpha\beta\mu} \frac{(\gamma \cdot p_u +
M_\Theta)} {u - M^2_\Theta}  \right. \nonumber \\
&& \qquad \qquad \;\; \left. \cdot\, S_{\beta\nu} \,\cdot\,
\gamma^{\nu\sigma\rho} \,
\frac{(p_K)_\sigma\cdot(p_u)_\rho}{M_\Theta^2} N  \right. \nonumber \\
&\;\;\;+& \; \left. (t-m^2_K) \, \cdot \bar{\Theta}_{\alpha} \,
\gamma^{\alpha\mu\nu} \, \frac{(F_\Theta(u) \,\cdot\, p_K \,+\,
F_K(t) \,\cdot\, p_\Theta)_\nu}{M_\Theta} N  \right. \nonumber \\
&\;\;\;+& \; \left. 2 p_K^\mu \,\cdot\, ({\hat F}(s,t,u) - F_K(t))
\,\cdot\, p_K^\alpha \,\cdot\, \bar{\Theta} \, N \right. \nonumber \\
&\;\;\;-& \; \left.  \left(\frac{t - m_K^2}{u - m_\Theta^2}\right)
\,\cdot\, 2 p_\Theta^\mu \,\cdot\, \{{\hat F}(s,t,u) - F_\Theta(u)\}
\cdot\, p_K^\alpha \,\cdot\, \bar{\Theta} \, N \right] \, , \nonumber \\
\label{eq:amplkaon_gankm_pos3par}
\end{eqnarray}

\begin{eqnarray}
{\mathcal M}_{K} \left(\gamma n \to K^{-} \Theta^+ : J^P_\Theta =
\frac{3}{2}^- \right)  &=&\,
\frac{-e \, g_{K N \Theta}}{m_K} \, \cdot \,
\mathcal{P}^{K}_{Regge}(s, t) \cdot \varepsilon_\mu (q, \lambda) \nonumber
\\
&\times&
\left[F_K(t) \,\cdot\, (2 p_K - q)^{\mu} \,\cdot\, (p_K - q)^{\alpha}
\,\cdot\, \bar{\Theta}_{\alpha} \, \gamma^5 \, N \right. \nonumber \\
&\;\;\;-& \; \left. F_\Theta(u) \,\cdot\, (t-m^2_K) \, \cdot
\bar{\Theta}_{\alpha} \, \gamma^{\alpha\beta\mu} \frac{(\gamma \cdot p_u +
M_\Theta)} {u - M^2_\Theta}  \right. \nonumber \\
&& \qquad \qquad \;\; \left. \cdot\, S_{\beta\nu} \,\cdot\,
\gamma^{\nu\sigma\rho} \, \frac{(p_K)_\sigma\cdot(p_u)_\rho}{M_\Theta^2} \,
\gamma^5N  \right. \nonumber \\
&\;\;\;+& \; \left. (t-m^2_K) \, \cdot \bar{\Theta}_{\alpha} \,
\gamma^{\alpha\mu\nu} \, \frac{(F_\Theta(u) \,\cdot\, p_K \,+\, F_K(t)
\,\cdot\, p_\Theta)_\nu}{M_\Theta} \, \gamma^5 N \right. \nonumber \\
&\;\;\;+& \; \left. 2 p_K^\mu \,\cdot\, ({\hat F}(s,t,u) - F_K(t))
\,\cdot\, p_K^\alpha \,\cdot\, \bar{\Theta} \, \gamma^5 \, N \right.
\nonumber \\
&\;\;\;-& \; \left.  \left(\frac{t - m_K^2}{u - m_\Theta^2}\right)
\,\cdot\, 2 p_\Theta^\mu \,\cdot\, \{{\hat F}(s,t,u) - F_\Theta(u)\}
\cdot\, p_K^\alpha \,\cdot\, \bar{\Theta} \, \gamma^5 \, N \right] \, ,
\nonumber \\
\label{eq:amplkaon_gankm_neg3par}
\end{eqnarray}

with
\begin{eqnarray}
S_{\beta\nu} = g_{\beta\nu} - \frac{\gamma_\beta \, \gamma_\nu}{3}  -
\frac{\left(\gamma_\beta \, (p_u)_\nu - \gamma_\nu \,
(p_u)_\beta\right)} {3M_\Theta}  - \frac{2\left((p_u)_\beta
(p_u)_\nu\right)}{3M_\Theta}  \, . \nonumber
\label{eq:theta_propagator}
\end{eqnarray}

Similar to the spin-1/2 case, the form factors $F_K(t)$ and $F_\Theta(u)$
have to be added to take into account the change of coupling constant
$g_{KN\Theta}$ away from the pole position.
In the calculation, we choose the same forms of $F_K(t)$
and $F_\Theta(u)$ for spin-3/2 as for the spin-1/2 case.   For the
spin-3/2 case we have to add the third, fourth and fifth terms in
Eqs.~(\ref{eq:amplkaon_gankm_pos3par},\ref{eq:amplkaon_gankm_neg3par}),
which originate from the contact diagrams to the $t$- and $u$-channel
process to preserve gauge invariant in the amplitude.  In the Regge
approach, this gauge restoring term is reggeized in the same way
as for the $t$-channel process, as discussed before for the spin-1/2 case.
We also choose the $\hat F(s,t, u)$ to have the same form as for the spin-1/2
case.

\subsection{$K^*$ Regge exchange}

We next consider the $K^*(892)$ exchange processes to both the
$\gamma n \to K^- \Theta^+$ and $\gamma p \to \bar K^0 \Theta^+$
reactions as shown in Fig.~\ref{fig:diag}. Note that
for the $\gamma p \to \bar K^0 \Theta^+$ reaction, $K$ exchange is
not possible as the real photon does not couple to the neutral kaon.
Therefore, we expect $K^*$ exchange to be the dominant $t$-channel
mechanism for the $\gamma p \to \bar K^0 \Theta^+$ reaction.
\newline
\indent
The construction of the reggeized amplitude for the $K^*$-exchange
processes, amounts to replace the $K^*$ pole by a
`Regge propagator' function $\mathcal{P}^{K^*}_{Regge}(s,t)$~:
\begin{eqnarray}
\frac{1}{t-m_{K^*}^2}\Longrightarrow
\mathcal{P}^{K^*}_{Regge}=\left({\frac{s}{s_0}}\right)^{\alpha_{K^*}(t)-1}
\frac{\pi \alpha_{K^*}'}{\sin(\pi \alpha_{K^*}(t))}\;
\frac{\mathcal{S}_K^*+e^{-i\pi \alpha_{K^*}(t)}}{2}\;
\frac{1}{\Gamma(\alpha_{K^*}(t))}\;,
\label{eq:kstarregge}
\end{eqnarray}
For the $K^*(892)$, we use a standard linear trajectory~:
\begin{eqnarray}
\alpha_{K^*}(t)= 0.25 + \alpha'_{K^*} \,t \, ,
\label{eqkkst}
\end{eqnarray}
where $\alpha'_{K^*} = 0.83$~GeV$^{-2}$.
Furthermore, we also consider a degenerate trajectory for $K^*$ leading to a
Regge propagator with rotating phase in Eq.~(\ref{eq:kstarregge}),
in line with our previous findings in the analysis
of strangeness photoproduction reactions at high energies~\cite{GLV97}.
\newline
\indent
To evaluate the $K^*$ processes in Fig.~\ref{fig:diag},
we next have to specify the $K^* N \Theta^+$ vertex function.
If the spin parity assignment of the $\Theta^+$ resonance is given by
$J^P = \frac{1}{2}^+$,
the $K^* N \Theta^+$ vertex can be written as~:
\begin{eqnarray}
\mathcal{L}_{K^* N \Theta} \,=\, f_{K^* N \Theta} \, \bar \Theta \,
\left[ \frac{i \, \sigma_{\mu \nu} \, p_{K^*}^\nu}{M_N + M_\Theta}\right] \, N
\, \cdot \, V^{\mu}(p_{K^*}) \,+\, \mathrm{h.c.} \; ,
\label{eq:kstar_pospar}
\end{eqnarray}
where $V^{\mu}(p_{K^*})$ is the polarization vector of the $K^*$ meson.
If the spin parity assignment of the $\Theta^+$ resonance is given by
$J^P = \frac{1}{2}^-$,
the $K^* N \Theta^+$ vertex is given by~:
\begin{eqnarray}
\mathcal{L}_{K^* N \Theta} \,=\, -i \, f_{K^* N \Theta} \, \bar \Theta \,
\left[ \frac{i \, \sigma_{\mu \nu} \, p_{K^*}^\nu}{M_N +
M_\Theta}\right] \, \gamma_5 \, N
\, \cdot \, V^{\mu}(p_{K^*}) \,+\, \mathrm{h.c.} \; .
\label{eq:kstar_negpar}
\end{eqnarray}
Using $SU(3)$ symmetry for the vector meson couplings
within the baryon octet and between the baryon octet and antidecuplet,
one can express~:
\begin{eqnarray}
g_{\rho^0 p p} + f_{\rho^0 p p}
&=& \frac{7}{10} \left( V_0 + \frac{1}{2} V_1 \right) + \frac{1}{20} V_2 ,
\label{eq:vm1} \\
g_{\omega p p} + f_{\omega p p}
&=& \frac{1}{10} \left( V_0 + \frac{1}{2} V_1 \right) + \frac{23}{20} V_2 ,
\label{eq:vm2} \\
g_{\phi p p} + f_{\phi p p}
&=& -\frac{1}{10} \left( V_0 + \frac{1}{2} V_1 \right) + \frac{7}{20} V_2 ,
\label{eq:vm3} \\
f_{K^{* 0} \Theta^+ p} &=& \frac{3}{\sqrt{30}}
\left( V_0 - V_1 - \frac{1}{2} V_2 \right) ,
\label{eq:vm4}
\end{eqnarray}
where $g_{VNN} (f_{VNN})$ are the vector (tensor) coupling constants
respectively.  First we use the fact that $g_{\phi p p} + f_{\phi p p}
\simeq 0$~\cite{Dumbrajs}, and solve for V$_2$ in
Eq.~(\ref{eq:vm3}).  Substituting this into Eqs.~(\ref{eq:vm1}) and
(\ref{eq:vm4}), we can express the $K^* N \Theta$ coupling as~:
\begin{eqnarray}
f_{K^{* 0} p \Theta^+} \,=\, (g_{\rho^0 pp} + f_{\rho^0 pp}) \,
\frac{3 \sqrt{3}}{\sqrt{10}} \, \frac{4/5 - r}{r + 2} \, ,
\label{eq:kstarntheta}
\end{eqnarray}
where $r$ is defined as $r \equiv V_1 / V_0$.  By fixing the ratio $r$ to
its value obtained in the chiral quark soliton model~\cite{DPP97}
: $r \simeq 0.35$ and using the value $g_{\rho^0 p p} + f_{\rho^0 p p}
\simeq 18.7$~\cite{Hoehler}, Eq.~(\ref{eq:kstarntheta}) then
leads to the coupling $f_{K^{*} N \Theta} =  \,\simeq\, 5.9$.
Note that the value of $f_{K^{* 0} p \Theta^+}$ is very sensitive
to the value of the parameter $r$, as there is a strong
cancellation of various contributions.
The case of 1~MeV width of $\Theta^+$
indicates that this cancellation can be very deep.
The above mentioned value
obtained in the particular dynamical model is subjected to large
theoretical uncertainties which result in big spread in values of
the coupling constant.
Here we assume that the parameter $r$ is close to the value of its
counterpart for the axial transitions which corresponds with a 
width of $\Theta^+$ of about 1~MeV.
This assumption
%
%
yields~:
\begin{eqnarray}
f_{K^{*} N \Theta} \,\equiv\,f_{K^{* 0} p \Theta^+}
\,\simeq\, 1.1 \, ,
\label{eq:kstarntheta2}
\end{eqnarray}
which will be used for the coupling constant entering in the vertices of
Eqs.~(\ref{eq:kstar_pospar},\ref{eq:kstar_negpar}) for
both parities of the $\Theta^+$.
\newline
\indent
The reggeized $K^*$ exchange amplitudes for both parities of the
$\Theta^+$ are then given by~:
\begin{eqnarray}
{\mathcal M}_{K^*} \left(\gamma p \to \bar K^{0} \Theta^+ :
J^P_\Theta = \frac{1}{2}^+ \right) &=&\,
i \, e\, f_{K^{* 0} K^0 \gamma} \,\, f_{K^{*} N \Theta}
\, \cdot \, \mathcal{P}^{K^*}_{Regge}(s,t) \, \cdot \,
\varepsilon^\mu(q, \lambda) \, \nonumber \\
&\times& \varepsilon_{\mu \nu \lambda \alpha }
\,q^\nu \,(q - p_K)^\lambda \, \bar \Theta \,
\left[ \, \frac{i \, \sigma^{\alpha \beta} \, (q - p_{K})_\beta}
{M_N + M_\Theta}\right] \, N \, ,
\label{eq:amplkstar_gapko_pospar} \\
{\mathcal M}_{K^*} \left(\gamma p \to \bar K^{0} \Theta^+ :
J^P_\Theta = \frac{1}{2}^- \right) &=&\,
e\, f_{K^{* 0} K^0 \gamma} \,\, f_{K^{*} N \Theta}
\, \cdot \, \mathcal{P}^{K^*}_{Regge}(s,t) \, \cdot \,
\varepsilon^\mu(q, \lambda) \, \nonumber \\
&\times& \varepsilon_{\mu \nu \lambda \alpha }
\,q^\nu \,(q - p_K)^\lambda \, \bar \Theta \,
\left[ \, \frac{i \, \sigma^{\alpha \beta} \, (q - p_{K})_\beta}
{M_N + M_\Theta}\right] \, \gamma_5 \, N \, ,
\label{eq:amplkstar_gapko_negpar}
\end{eqnarray}
and analogous formulas hold for the $K^*$ contribution to the
$\gamma n \to K^- \Theta^+$ reaction.
In Eqs.~(\ref{eq:amplkstar_gapko_pospar},\ref{eq:amplkstar_gapko_negpar}),
the electromagnetic coupling $f_{K^{*} K \gamma}$ can be extracted
from the radiative decay widths
$\Gamma_{K^{* 0}\to K^0 \gamma} = 0.117$ MeV, and
$\Gamma_{K^{* -}\to K^- \gamma} = 0.05$ MeV, yielding~\cite{GLV97}~:
\begin{eqnarray}
f_{K^{* 0} K^0 \gamma}  \,&=&\, 1.28 \;\; \mathrm{GeV}^{-1} \, , \\
f_{K^{* -} K^- \gamma}  \,&=&\, 0.84 \;\; \mathrm{GeV}^{-1} \, .
\end{eqnarray}
Note that the $K^*$ $t$-channel exchange amplitudes
of Eqs.~(\ref{eq:amplkstar_gapko_pospar},\ref{eq:amplkstar_gapko_negpar})
are gauge invariant by themselves.

\section{Results}

\begin{figure}[ht]
\includegraphics[width=14cm]{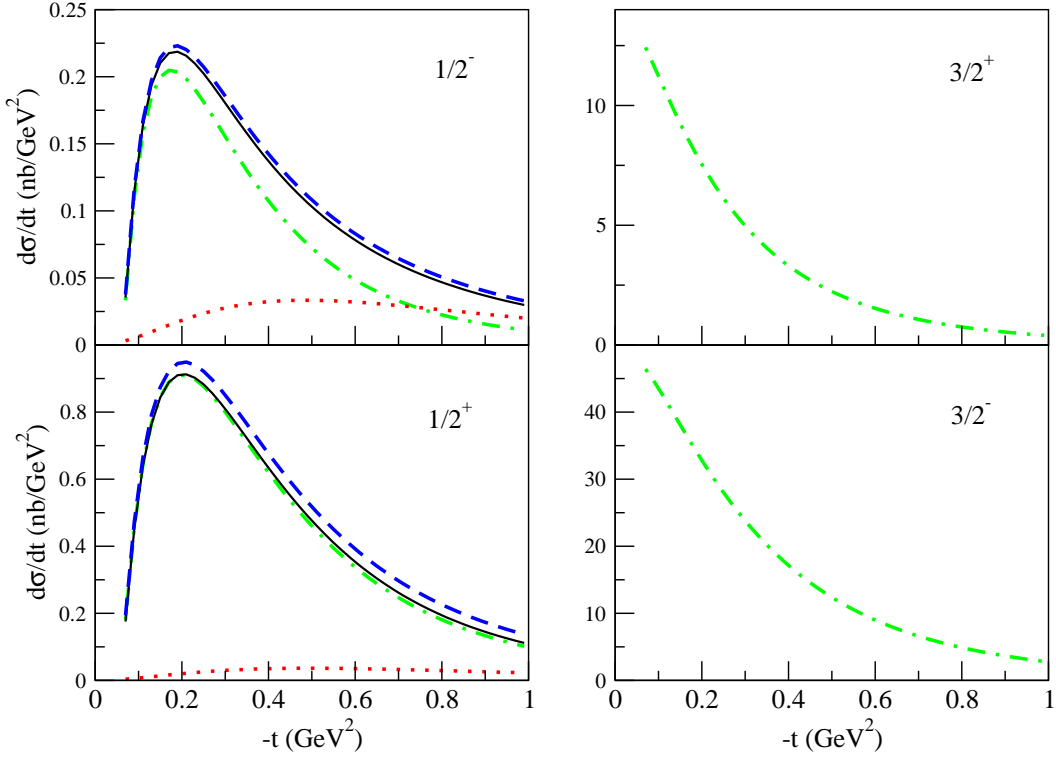}
\caption{Regge model predictions for the $\gamma n \to K^- \Theta^+$
cross section at $E_\gamma = 4$~GeV
for different spin parity assignments of the $\Theta^+$.
Dashed-dotted curves : gauge invariant $K$ Regge exchange;
dotted curves : $K^*$ Regge exchange (for the cases of $1/2^\pm$).
To account for the range of uncertainty
in the $K^* N \Theta$ coupling, we display the result for $K + K^*$ exchange
for two values of the $K^* N \Theta$ coupling constant :
$f_{K^* N \Theta} = +1.1$ (solid curves), and
$f_{K^* N \Theta} = -1.1$ (dashed curves),
corresponding with a width $\Gamma_{\Theta \to K N} = 1$~MeV.}
\label{fig:gndiff}
\end{figure}

\begin{figure}[ht]
\includegraphics[width=14cm]{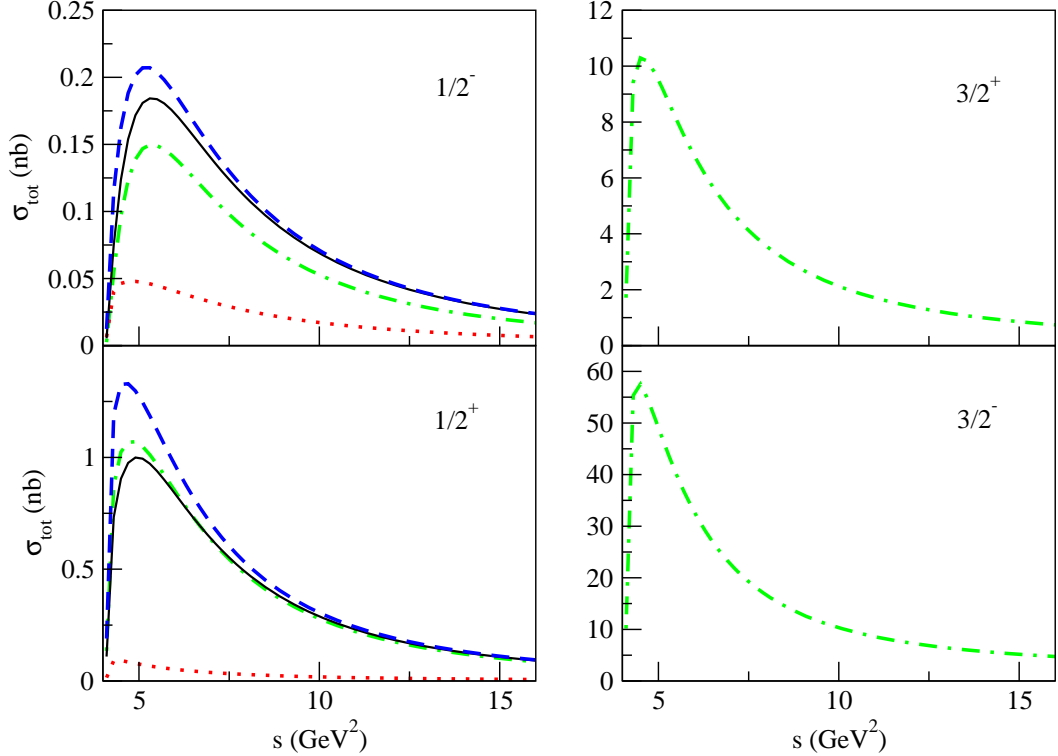}
\caption{Regge model predictions for the $\gamma n \to K^- \Theta^+$
total cross sections
for different spin-parity assignments of the $\Theta^+$ resonance.
Curve conventions as in Fig.~\ref{fig:gndiff}.}
\label{fig:gntot}
\end{figure}

\begin{figure}[ht]
\includegraphics[width=14cm]{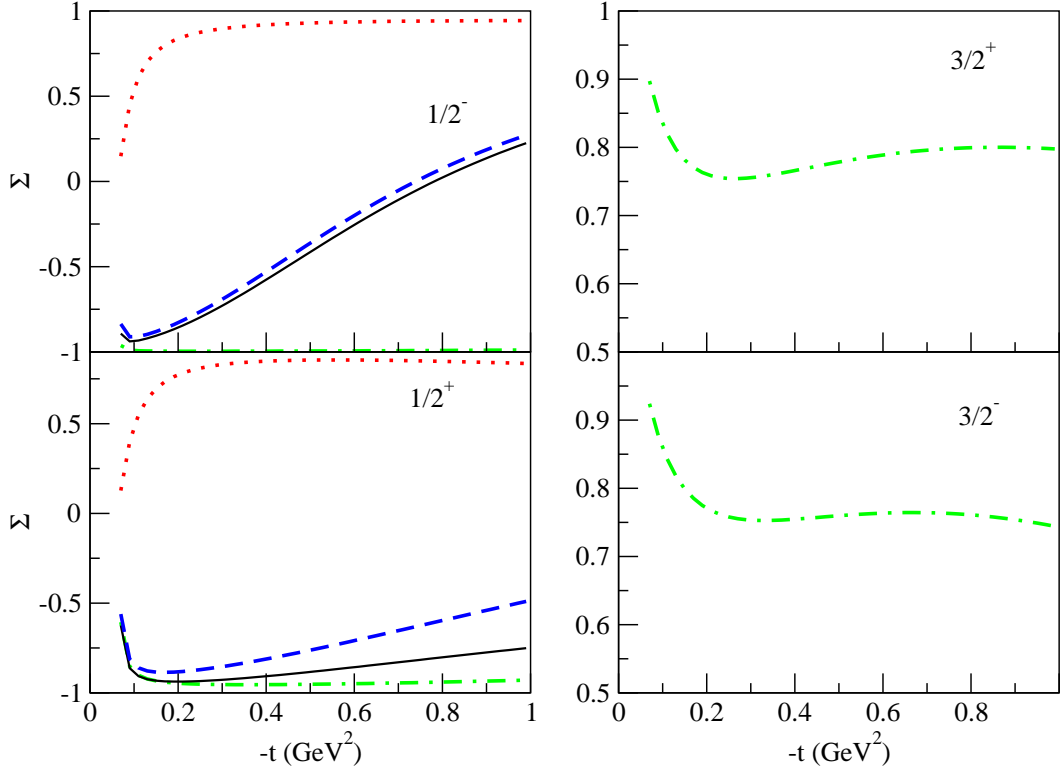}
\caption{Regge model predictions for the $\gamma n \to K^- \Theta^+$
photon asymmetry at $E_\gamma = 4$~GeV
for different spin parity assignments of the $\Theta^+$.
Curve conventions as in Fig.~\ref{fig:gndiff}.}
\label{fig:gngaasy}
\end{figure}

In Fig.~\ref{fig:gndiff}, we show our results for the differential
cross section for the $\gamma n \to K^- \Theta^+$ reaction
for the cases of spin-parity assignments $1/2^\pm$ and
$3/2^\pm$ of the $\Theta^+$.
Comparing the cross sections for $K$ Regge exchange for the cases of
$J^P =  1/2^\pm$, one notices
that the $K N \Theta$ coupling constant for the case of a
negative parity $\Theta^+$ is a factor 7 smaller than for the
case of a positive parity $\Theta^+$.
If there were only $t$-channel $K$ exchange, this would result in a ratio of
about a factor 50 for the cross sections of positive parity compared to
negative parity $\Theta^+$ photoproduction
for the case of a $\Theta^+$ of $J = 1/2$.
The cross section for $J^P_{\Theta} = 1/2^-$ gets enhanced
though through the $u$-channel $\Theta^+$ process
and the contact diagrams, which are required to make the $t$-channel
charged kaon exchange process gauge-invariant.
They are also responsible for the
pronounced peak structure in the differential cross section at
low values of $-t$.
For the cases of $J^P_{\Theta} = 3/2^\pm$, the situation is reversed, as
the $K N \Theta$ coupling constant is about a factor 7 larger for
the case of $3/2^-$ compared with the case of $3/2^+$
when using a same width for the $\Theta^+$.
Taking into account the $u$-channel and contact diagrams then yields cross
sections for the case of $3/2^-$ which are about a factor
of 4 larger than for the case of $3/2^+$.

In Fig.~\ref{fig:gndiff}, we also show our estimates for
the $K^*$ exchange process for the spin-parity assignments $1/2^\pm$.
To show the range of uncertainty arising
from the $K^* N \Theta$ coupling, we display our
results for two values : $f_{K^* N \Theta} = +1.1$
and $f_{K^* N \Theta} = -1.1$.
The value $f_{K^* N \Theta} = +1.1$ is obtained by
rescaling the chiral quark soliton model
coupling for the case $J^P_{\Theta} = 1/2^+$ by the same amount
as when rescaling the model value of $g_{K N \Theta}$
to correspond to a width of 1 MeV.
One notices from Fig.~\ref{fig:gndiff} that for a $K^* N \Theta$ coupling
within this range, the resulting $K^*$ Regge
exchange process yields only a very small contribution
to the cross section compared with
the gauge-invariant $K$ exchange, in particular for the case of $1/2^+$.
Furthermore, in the forward direction the $K^*$ exchange process
vanishes due to the momentum $(q - p_K)$ dependence
in the $\gamma K K^*$ vertex.
Such a behavior has been confirmed by data for
the $\gamma p \to K^+ \Sigma^0$ reaction which is dominated by $K^*$
exchange at large $s$ and small $-t$, see Ref.~\cite{GLV97}.
Analogously, the forward angular region ($-t << s$) for the
$\gamma n \to K^- \Theta^+$ reaction at high photon energy is dominated by
charged $K$ exchange.
At larger values of $-t$ the relative weight of $K^*$ versus $K$ exchange
increases.
Using the same values for the $K^* N \Theta$ coupling in case
$J^P_{\Theta} = 1/2^-$, the $K^*$ exchange contribution becomes comparable
to the gauge-invariant $K$ exchange for values around $-t \simeq 1$~GeV$^2$.

In Fig.~\ref{fig:gntot}, we show the corresponding total cross sections.
Using a width of 1 MeV for the $\Theta^+$,
the maximum value of the total cross sections can be seen to be around
1 nb for the case of $1/2^+$ and 0.2 nb for the case of $1/2^-$.
For the cases of $3/2^+$ ($3/2^-$) much larger cross sections of around
10 nb (55 nb) are obtained when using a same value of 1 MeV for
the $\Theta^+$ width.

A direct measure of the relative weight of $K$ versus $K^*$ exchange
processes can be obtained by the linear photon asymmetry, defined as~:
\begin{eqnarray}
\Sigma = \frac{\sigma_\perp - \sigma_\parallel}{\sigma_\perp + \sigma_\parallel} ,
\end{eqnarray}
where $\sigma_\parallel$ and $\sigma_\perp$ are the cross sections induced by
a linearly polarized photon beam with polarization vector lying in the
reaction plane (for $\sigma_\parallel$)
and perpendicular to the reaction plane (for $\sigma_\perp$) respectively.
At high $s$ and small $-t$ (with $-t << s$) the photon asymmetry for
a natural parity $t$-channel exchange (such as for the $K^*$)
approaches the value +1 (i.e. $\sigma_\perp$ dominates),
whereas the photon asymmetry for an unnatural parity
$t$-channel exchange process (such as for the $K$) yields
the value -1 (i.e. $\sigma_\parallel$ dominates). The $u$-channel process
and contact diagram,
which are needed to make the $t$-channel $K$ exchange gauge invariant are
responsible for the deviation of the photon asymmetry from the
value of -1, as is seen on Fig.~\ref{fig:gngaasy}.
As the $\gamma n \to K^- \Theta^+$ process at large $s$ and low $-t$
is dominated by $K$ exchange, one sees from Fig.~\ref{fig:gngaasy}
that the photon asymmetry rises sharply, at small $-t$,
to a large negative value for the cases of $1/2^\pm$.
At larger values of $-t$ (for $-t \geq 0.2$~GeV$^2$), one
sees from Fig.~\ref{fig:gngaasy} that the influence of the $K^*$ exchange
in the photon asymmetry shows up, in particular for the case of $1/2^-$.
The photon asymmetry seems therefore to be
a very promising signature to distinguish between the
$J^P_{\Theta} = 1/2^+$ and $J^P_{\Theta} = 1/2^-$ cases.

\begin{figure}[ht]
\vspace{1.5cm}
\includegraphics[width=8cm]{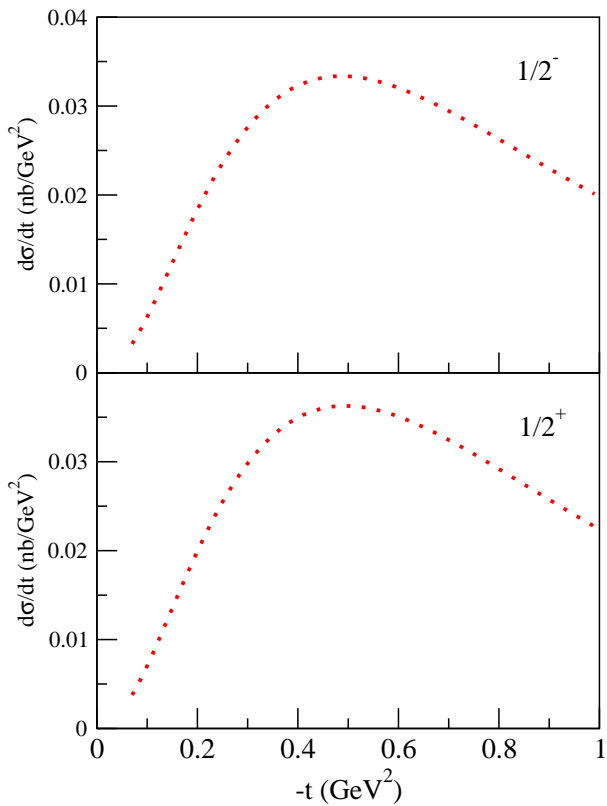}
\caption{Regge model predictions for the $\gamma p \to \bar K^0 \Theta^+$
cross section at $E_\gamma = 4$~GeV
for different spin-parity assignments of the $\Theta^+$.
Dotted curves : $K^*$ Regge exchange. }
\label{fig:gpdiff}
\end{figure}

\begin{figure}[ht]
\includegraphics[width=9cm]{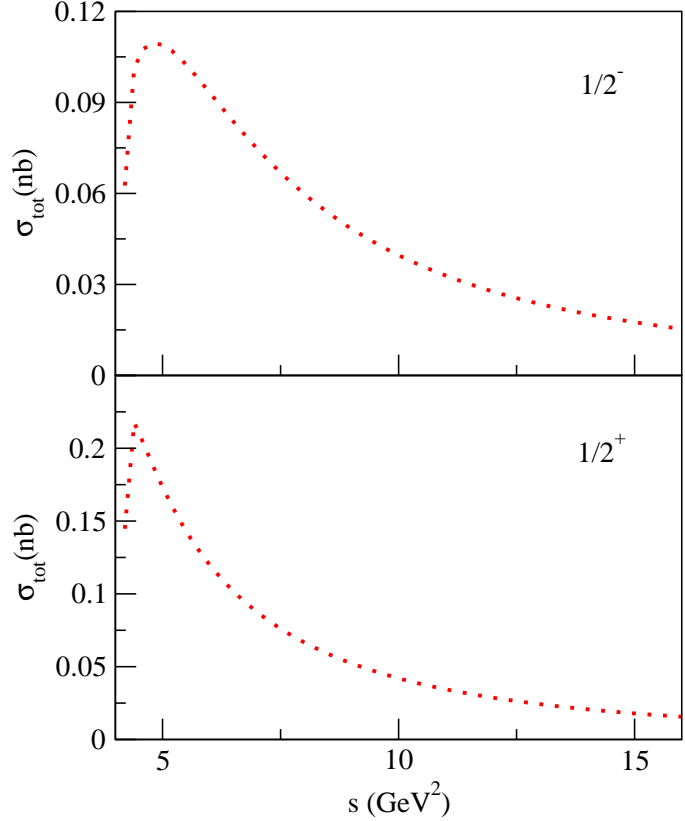}
\caption{Regge model predictions for the $\gamma p \to \bar K^0 \Theta^+$
total cross sections
for different spin-parity assignments of the $\Theta^+$ resonance.
Dotted curves : $K^*$ Regge exchange. }
\label{fig:gptot}
\end{figure}

\begin{figure}[ht]
\includegraphics[width=8cm]{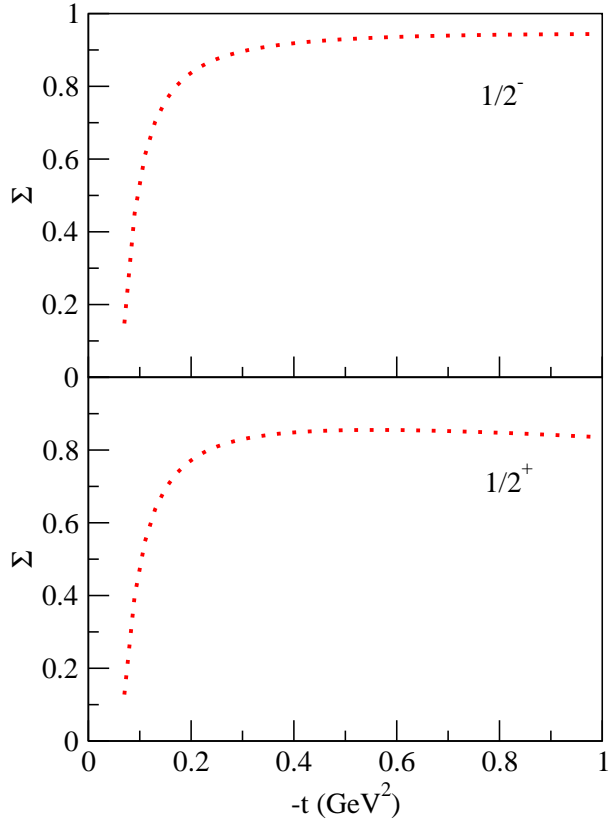}
\caption{Regge model predictions for the $\gamma p \to \bar K^0 \Theta^+$
photon asymmetry at $E_\gamma = 4$~GeV
for different spin-parity assignments of the $\Theta^+$.
Dotted curves : $K^*$ Regge exchange.}
\label{fig:gpgaasy}
\end{figure}

In Fig.~\ref{fig:gpdiff}, we show the corresponding observables for the
$\gamma p \to \bar K^0 \Theta^+$. For the neutral kaon production
reaction, the $t$-channel $K$ exchange is absent, and the dominant
$t$-channel mechanism is $K^*$ exchange. One therefore sees from
Figs.~\ref{fig:gpdiff},\ref{fig:gpgaasy}
that the observables carry the signatures of
a $K^*$ dominated process, i.e. a differential cross section which vanishes
in the low $t$ region and a photon asymmetry which reaches large positive
values. By comparing the processes on the neutron (in Fig.~\ref{fig:gntot})
and on the proton (in Fig.~\ref{fig:gptot}), one notices
that the absence of the $K$ exchange mechanism
yields cross sections on the proton that are about a factor 5 to 10
smaller than their counterparts on the neutron for the cases of
$J^P_{\Theta} = 1/2^\pm$.

\begin{figure}[ht]
\includegraphics[width=9cm]{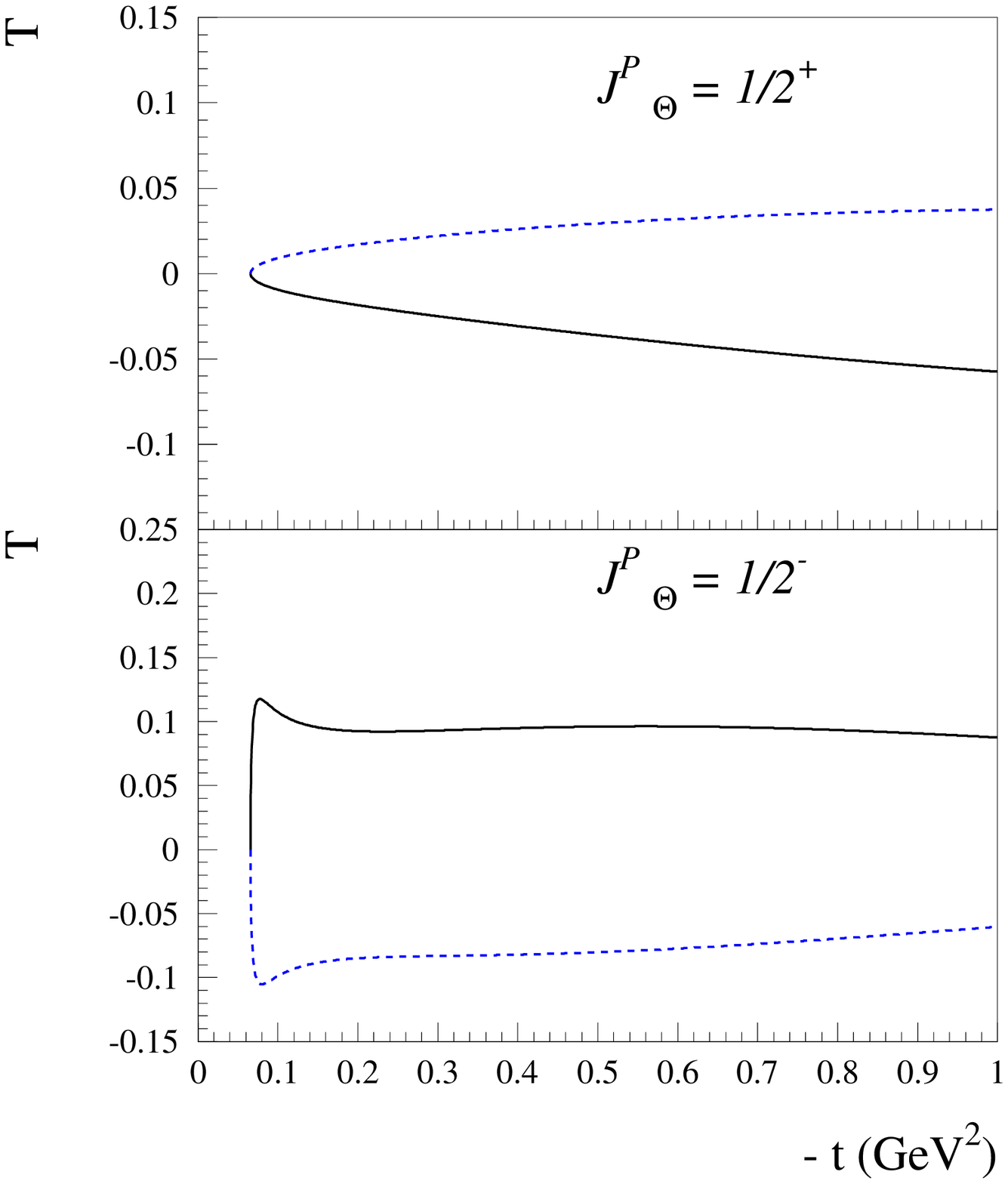}
\caption{Regge model predictions for the target
single spin asymmetry $T$ for the
$\gamma n \to K^- \Theta^+$ reaction
for both possible parities of the $\Theta^+$ resonance
(for $J = 1/2$) at $E_\gamma = 4$~GeV.
The model calculations correspond with the $K + K^*$ Regge exchanges
for two values of the $K^* N \Theta$ coupling.
Solid curves : $f_{K^* N \Theta} = +1.1$,
dashed curves : $f_{K^* N \Theta} = -1.1$.}
\label{fig:target}
\end{figure}

\begin{figure}[htbp]
\includegraphics[width=9cm]{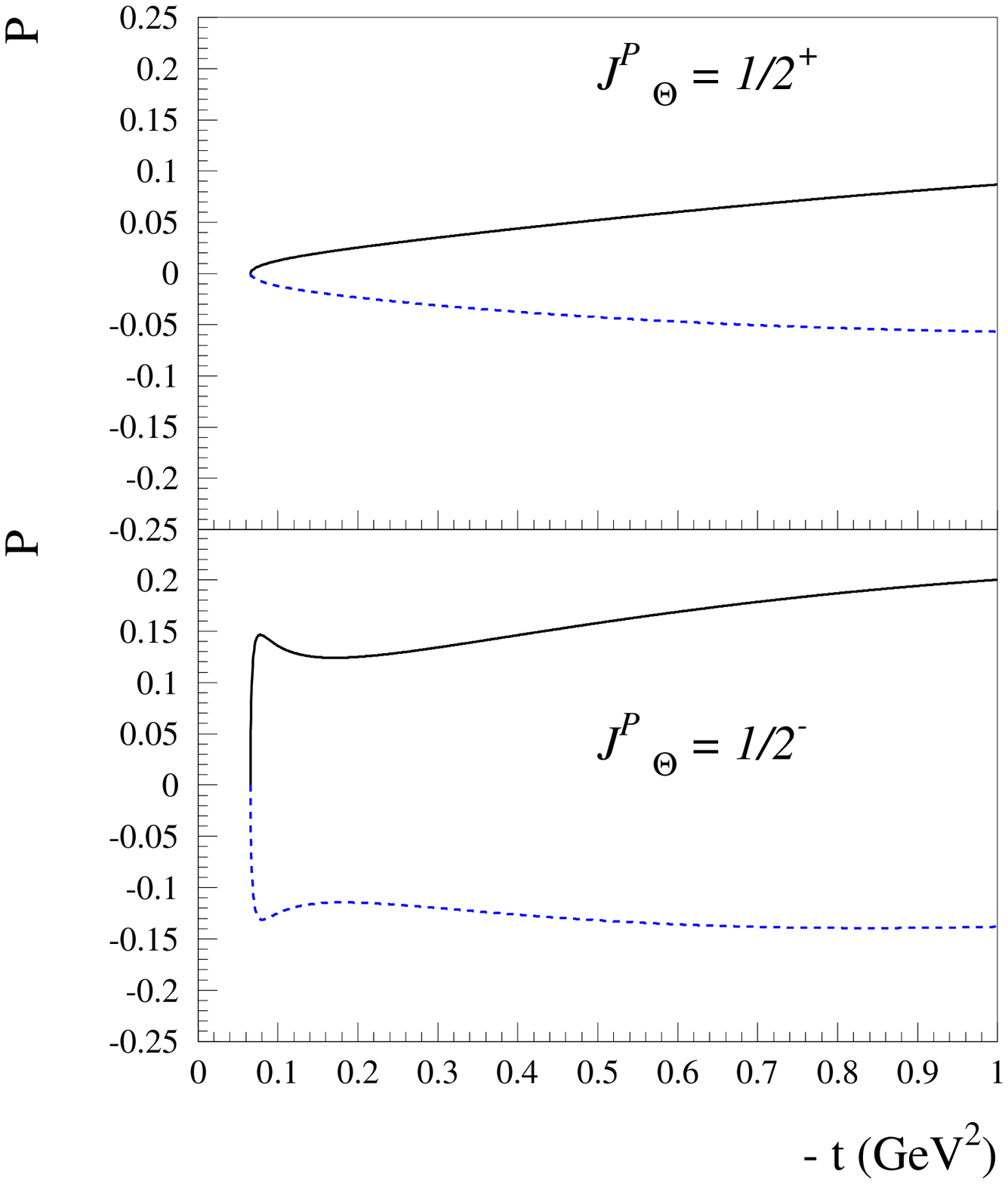}
\caption{Regge model predictions for the recoil
single spin asymmetry $P$ for the
$\gamma n \to K^- \Theta^+$ reaction
for both possible parities of the $\Theta^+$ resonance
(for $J = 1/2$) at $E_\gamma = 4$~GeV.
Curve conventions as in Fig.~\ref{fig:target}.}
\label{fig:recoil}
\end{figure}

We next study the sensitivity of single target or recoil
polarization observables for the $\gamma n \to K^- \Theta^+$ reaction,
to the spin-parity assignments
$J^P_{\Theta} = 1/2^+$ and $J^P_{\Theta} = 1/2^-$.
The single target spin asymmetry ($T$) is defined as~:
\begin{eqnarray}
T = \frac{\sigma_\Uparrow - \sigma_\Downarrow}{\sigma_\Uparrow + \sigma_\Downarrow} ,
\end{eqnarray}
where $\sigma_\Uparrow$ ($\sigma_\Downarrow$) are the cross sections
where the target spin is polarized along (opposite) to the
vector $\hat n \equiv ( \vec q \times \vec p_K ) / |\vec q \times \vec p_K|$,
normal to the reaction plane. The recoil spin asymmetry ($P$) is defined
in an analogous way, where the $\Theta^+$ has its spin polarized along
or opposite to the normal vector $\hat n$.
In Figs.~\ref{fig:target} and \ref{fig:recoil}, we compare the Regge model
results for $T$ and $P$ for both $\Theta^+$ spin-parity assignments
$1/2^\pm$.
We first notice that the observables $T$ and $P$ are proportional to
an imaginary part of the interference of two amplitudes.
Therefore, one only obtains a non-zero value for $T$ or $P$ when the
two interfering amplitudes exhibit a phase difference.
The $K$ exchange or the $K^*$ exchange processes by themselves give us
therefore a zero value for the asymmetries $T$ and $P$.
Their sum however leads to a non-zero value for $T$ and $P$ as shown in
Figs.~\ref{fig:target} and \ref{fig:recoil}, due to the phase difference
between the $K$ and $K^*$ Regge amplitudes.
We see from Figs.~\ref{fig:target} and \ref{fig:recoil} that for
$J^P_{\Theta} = 1/2^+$, $T$ and $P$ have an opposite sign.
On the other hand for the situation $J^P_{\Theta} = 1/2^-$,
$T$ and $P$ display the same sign.

\begin{figure}[ht]
\includegraphics[width=12cm]{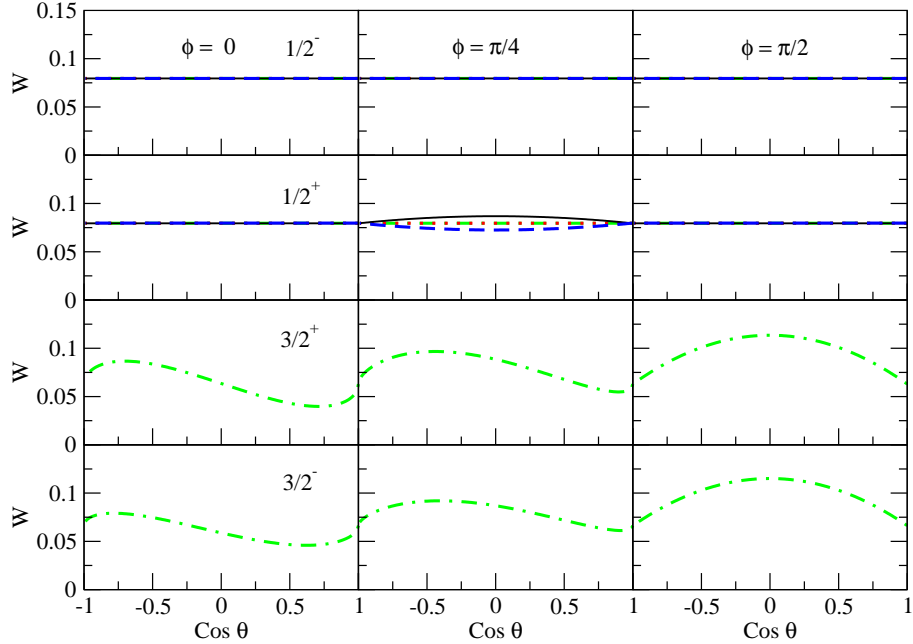}
\caption{Regge model predictions for the $\gamma n \to K^- \Theta^+$
angular distribution for an unpolarized photon (0)
for different spin parity assignments of the $\Theta^+$,
and for different decay angles ($\theta, \phi$),
defined in the rest frame of the $\Theta^+$.
Dashed-dotted curves : $K$ Regge exchange;
dotted curves : $K^*$ Regge exchange;
solid curves : $K + K^*$ Regge exchanges.}
\label{fig:monopole_ang_dist_gan_0.eps}
\end{figure}

\begin{figure}[ht]
\includegraphics[width=12cm]{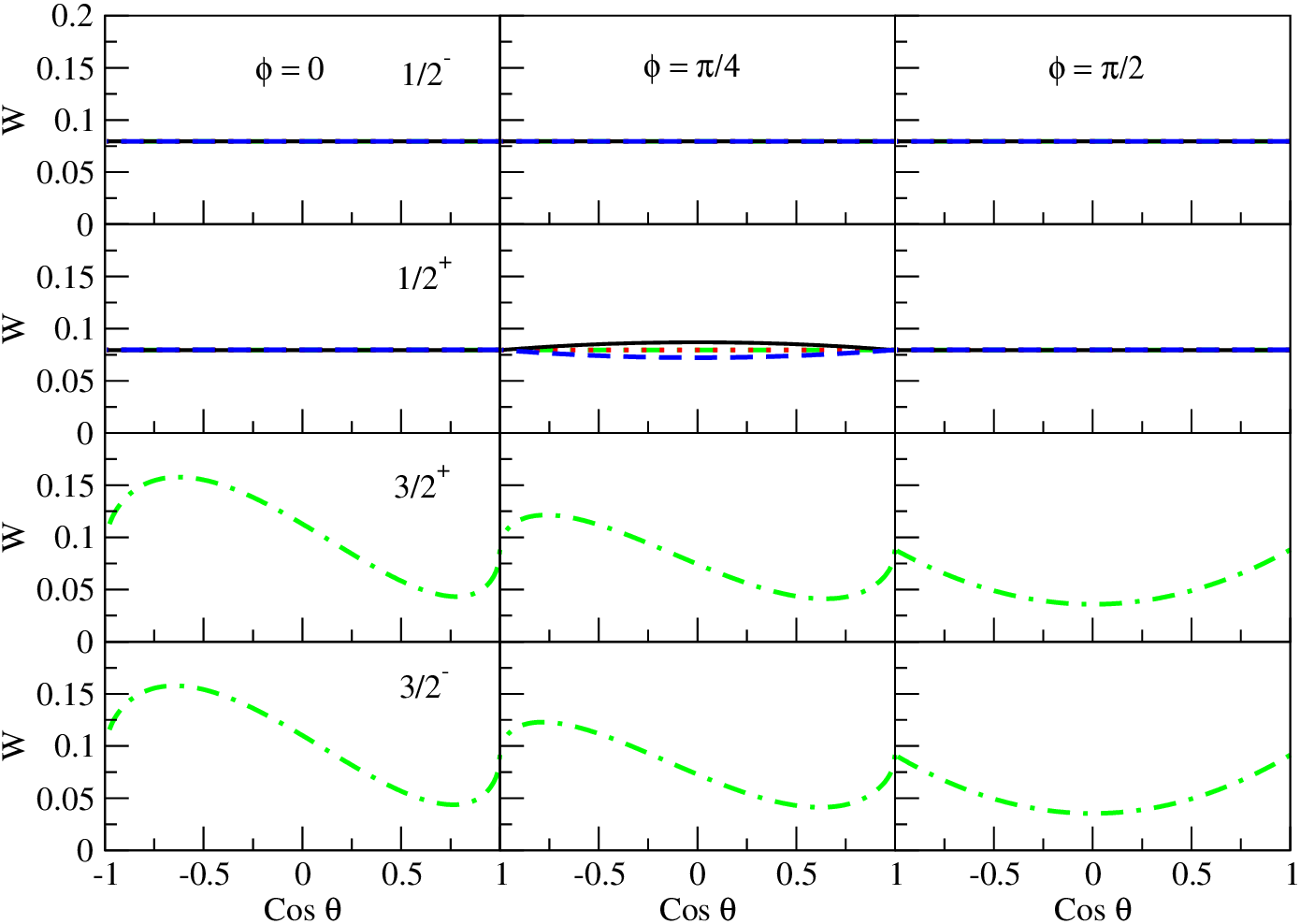}
\caption{Regge model predictions for the $\gamma n \to K^- \Theta^+$
angular distribution for a photon linearly polarized in the reaction plane
(x) for different spin parity assignments of the $\Theta^+$.
Curve conventions as in Fig.~\ref{fig:monopole_ang_dist_gan_0.eps}.}
\label{fig:monopole_ang_dist_gan_x.eps}
\end{figure}

\begin{figure}[ht]
\includegraphics[width=12cm]{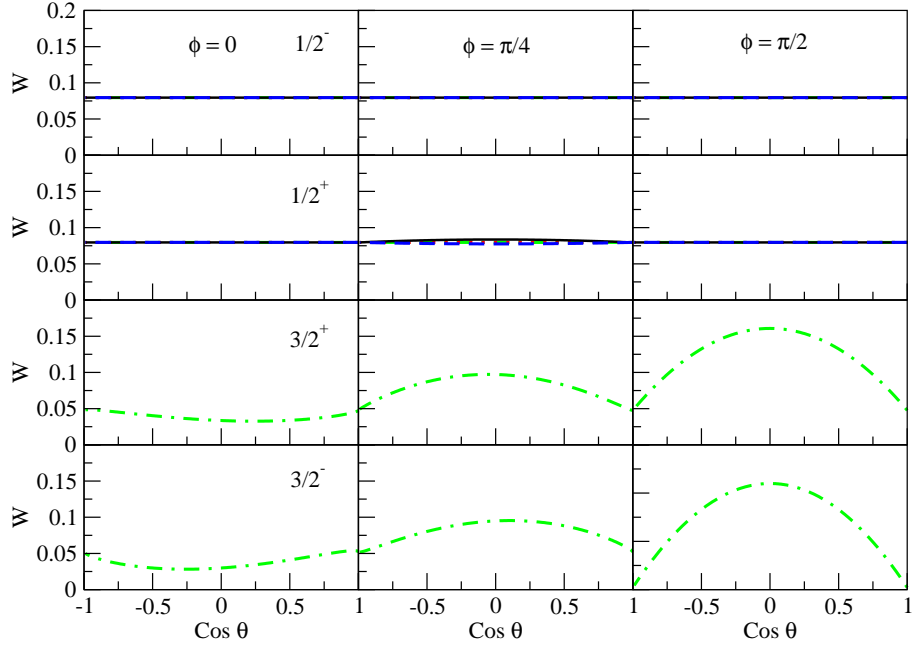}
\caption{Regge model predictions for the $\gamma n \to K^- \Theta^+$
angular distribution for a photon linearly polarized perpendicular to the
reaction plane (y) for different spin parity assignments of the
$\Theta^+$.
Curve conventions as in Fig.~\ref{fig:monopole_ang_dist_gan_0.eps}.}
\label{fig:monopole_ang_dist_gan_y.eps}
\end{figure}

\begin{figure}[ht]
\includegraphics[width=12cm]{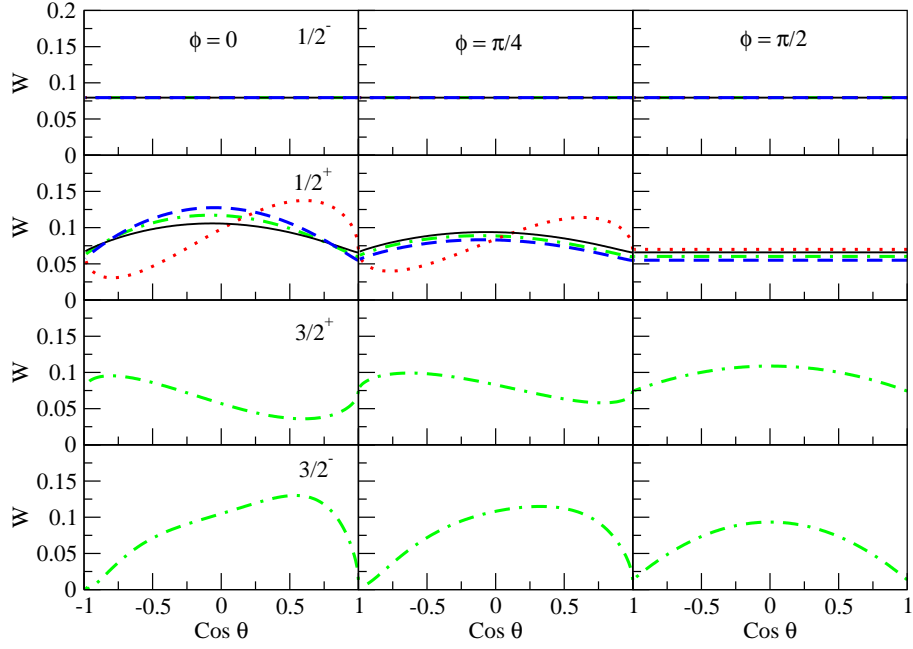}
\caption{Regge model predictions for the $\gamma n \to K^- \Theta^+$
angular distribution for a left-handed circularly polarized photon
(c, left) for different spin parity assignments of the $\Theta^+$.
Curve conventions as in Fig.~\ref{fig:monopole_ang_dist_gan_0.eps}.}
\label{fig:monopole_ang_dist_gan_c_left.eps}
\end{figure}

\begin{figure}[ht]
\includegraphics[width=12cm]{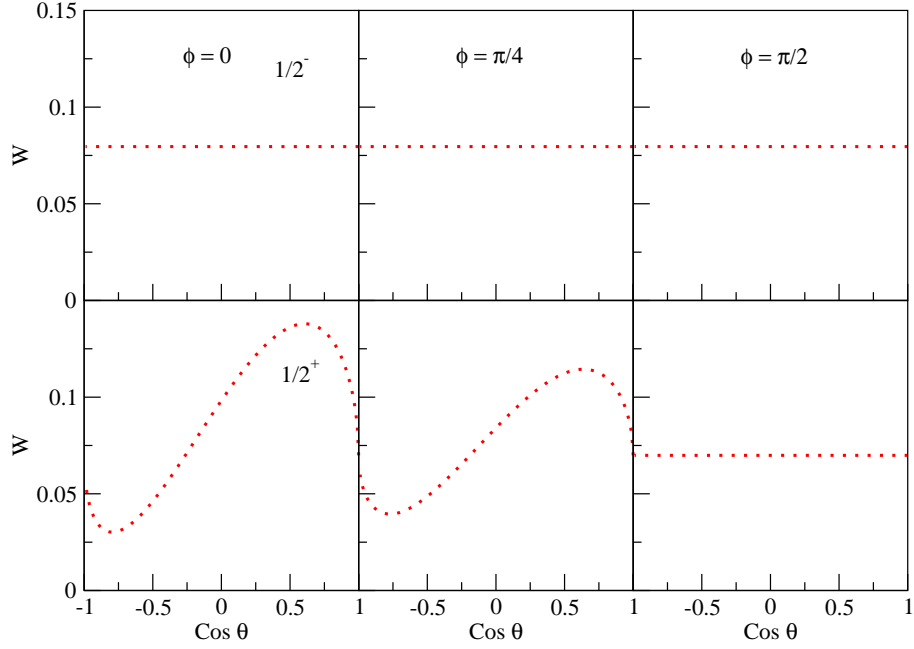}
\caption{Regge model predictions for the $\gamma p \to K^0 \Theta^+$
angular distribution (c, left) for different spin parity assignments of the
$\Theta^+$.
Dotted curves : $K^*$ Regge exchange.}
\label{fig:ang_dist_gap_c_left.eps}
\end{figure}

Besides the observables discussed above, where one integrates over all
possible final states for the $\Theta^+$ decay, one can also
observe the decay angular distributions of the $\Theta^+$.
They show a characteristic dependence on the spin and parity
of the final state.
In the appendix, we list the $\Theta^+$ decay angular distributions
for the spin-parity assignments $1/2^\pm$ and $3/2^\pm$.
\newline
\indent
We show the decay angular
distributions for different photon polarizations :
unpolarized (0) (Fig.~\ref{fig:monopole_ang_dist_gan_0.eps}),
linearly polarized in the reaction plane (x)
(Fig.~\ref{fig:monopole_ang_dist_gan_x.eps}),
linearly polarized perpendicular to the reaction plane (y)
(Fig.~\ref{fig:monopole_ang_dist_gan_y.eps}),
and left-handed circular polarization (c, left)
(Figs.~\ref{fig:monopole_ang_dist_gan_c_left.eps} and
\ref{fig:ang_dist_gap_c_left.eps}).
For the spin-parity assignments of $1/2^\pm$, one notices
that the decay angular distributions (0), (x) and (y) display a nearly flat
angular dependence. The decay angular distribution for a circularly
polarized photon (c, left) on the other hand, is flat for the case of $1/2^-$
but not uniform for $1/2^+$, allowing to distinguish between both parity
cases.
All decay angular distributions show characteristic angular dependences in
the case of $3/2^\pm$, which would be easily distinguishable from the
$1/2^\pm$ case.

\begin{figure}[ht]
\includegraphics[width=12cm]{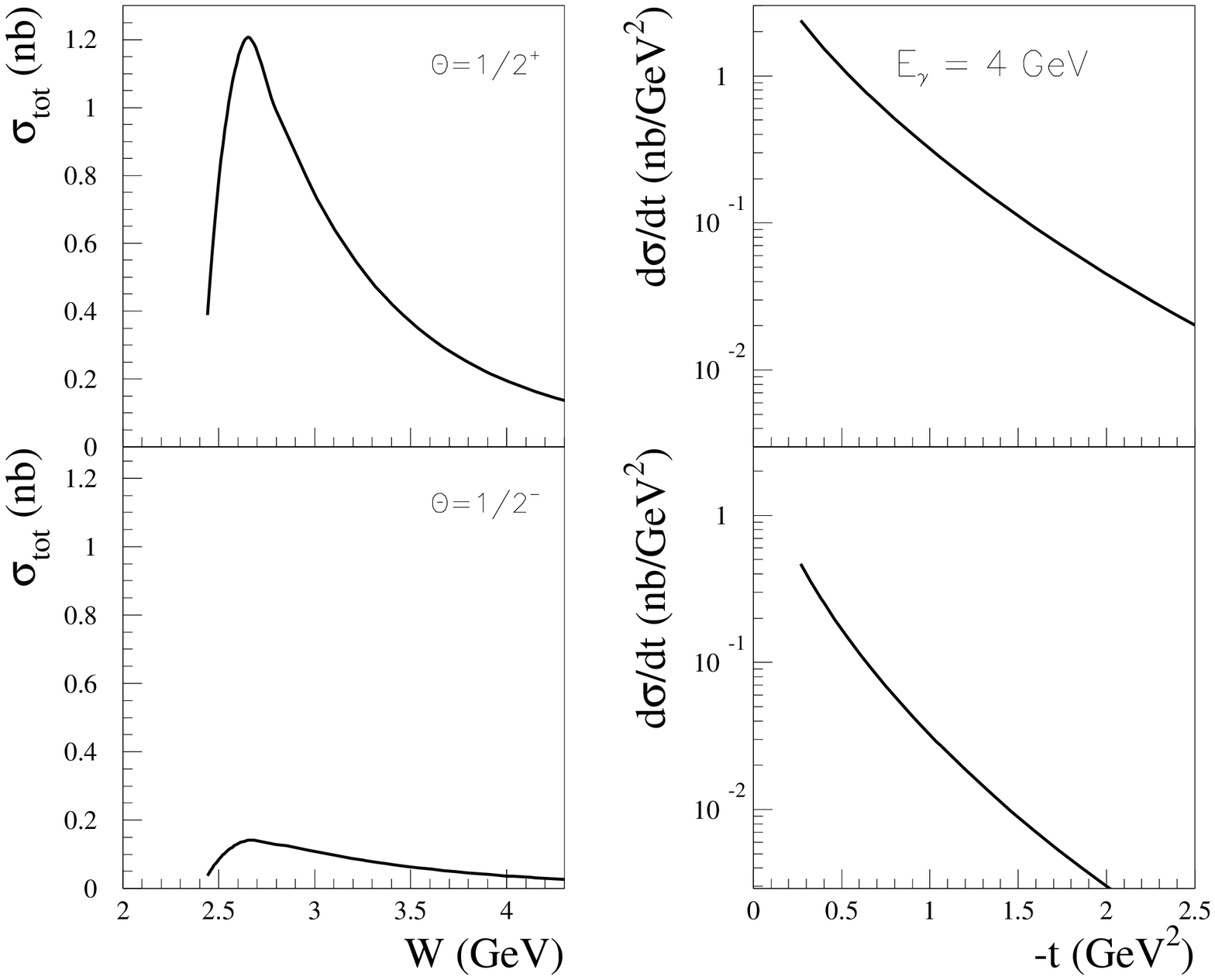}
\caption{Regge model predictions for the $\gamma p \to \bar K^{* 0} \Theta^+$
reaction for both possible parities of the $\Theta^+$ resonance.
Upper panels : positive parity case; lower panels : negative parity case.
Left panels : total cross section;
right panels : differential cross section at $E_\gamma = 4$~GeV.
The model calculation corresponds with $K^0$ Regge exchange.}
\label{fig:kst}
\end{figure}

In Fig.~\ref{fig:kst}, we also show our results for the
$\gamma p \to \bar K^{* 0} \Theta^+$ reaction. For this process,
the dominant $t$-channel exchange mechanism, at high $s$ and low $-t$,
is given by $K^0$ exchange as shown in Fig.~\ref{fig:diag} (lower right panel).
This yields to strong forwardly peaked angular distributions, as is seen
from Fig.~\ref{fig:kst}.  For a $\Theta^+$ width of 1 MeV, the
$1/2^+$ ($1/2^-$) total cross sections reach a maximum value of
1.2 nb (0.2 nb) respectively.

\section{Conclusions}

In this work, we studied the reaction mechanism
for the photoproduction of the
$\Theta^+(1540)$ resonance on the nucleon, through
$K$ and $K^*$ Regge exchanges.
Our estimates depend on only two parameters :
the $K N \Theta^+$ and $K^* N \Theta^+$ coupling constants.
The $K N \Theta^+$ coupling constant is directly related to the
$\Theta^+$ width. We determine the $K^* N \Theta^+$ coupling constant
by rescaling the value obtained from the chiral quark soliton model by the
same amount one has to rescale the $K N \Theta^+$ coupling to yield a given
value of the $\Theta^+$ width.

In the Regge model, which is assumed to be valid above {\it c.m.}
energies above 2 GeV, the $\Theta^+$ photoproduction cross sections
show a strong forward angular dependence.
We compared the size of the cross sections for the
$\gamma n \to K^- \Theta^+$ and $\gamma p \to \bar K^0 \Theta^+$ reactions, and
investigate their sensitivity to the spin-parity assignments $J^P =
\frac{1}{2}^\pm, \frac{3}{2}^\pm$ for the $\Theta^+$ resonance.
Using the Regge model, we
estimated the cross sections corresponding with a given
upper bound on the width of the $\Theta^+$.
Within this model, the cross sections on
the neutron were found to be around a factor 5 larger than the ones
on the proton, due to the presence of charged $K$ exchange for the reaction on
a neutron target.
For the case of spin-parity $J^P = \frac{1}{2}^+$, we found that
a $\Theta^+$ width of 1 MeV yields
$\gamma n \to K^- \Theta^+$ cross sections of around 1 nb,
and  $\gamma p \to \bar K^0 \Theta^+$  cross sections around 0.2 nb.
In the absence of a signal of the $\Theta^+$ in such reactions, our
estimates may be used to translate a given cross section upper limit into
an upper bound on the width of the $\Theta^+$.

Furthermore, we also estimated the photon asymmetry which was
found to display a pronounced sensitivity
to the parity of the $\Theta^+$. Provided the $\Theta^+$ can be produced,
the photon asymmetry would be a very promising observable
to help determining the quantum numbers of the $\Theta^+$ resonance.

\section*{Acknowledgments}
This work was supported by the U.S. Department of Energy under contract
DE-AC05-84ER40150.  H.J.K. is supported in part by the National Science
Foundation under grant No. PHY-0456525 and by a SURA Fellowship.  M.G. is
supported by the French Centre National de la Recherche Scientifique.
M.V.P. is supported by the Sofya Kovalevskaya Programme of the Alexader
von Humboldt Foundation.

\appendix

\section{Decay Angular Distribution}

After produced, the $\Theta^+$ decays into $\bar K^0 p$ or $K^+ n$
in 50\% ratio.  The angular distribution of the decay product
(Kaon) can be determined by:
\begin{eqnarray}
W(\theta,\phi) &=&
\sum_{s_f, s_f^\prime; s_\theta, s_\theta^\prime}
\hat{R}_{s_f,s_\theta} \rho_{s_\theta,s^\prime_\theta}(\Theta^+)
\hat{R}^*_{s_f^\prime,s_\theta^\prime} \nonumber\\
&=& \sum_{s_\theta, s_\theta^\prime}
\big\{ \hat{R}_{-\frac 12,s_\theta} \rho_{s_\theta,s^\prime_\theta}
\hat{R}^*_{-\frac 12,s_\theta^\prime}
+ \hat{R}_{\frac 12,s_\theta} \rho_{s_\theta,s^\prime_\theta}
\hat{R}^*_{-\frac 12,s_\theta^\prime} \nonumber\\
&& + \hat{R}_{\frac 12,s_\theta} \rho_{s_\theta,s^\prime_\theta}
\hat{R}^*_{\frac 12,s_\theta^\prime}
+ \hat{R}_{-\frac 12,s_\theta} \rho_{s_\theta,s^\prime_\theta}
\hat{R}^*_{\frac 12,s_\theta^\prime} \big\} \, ,
\end{eqnarray}
where the transition operator $\hat{R}_{s_f, s_\theta}$ is defined as
follow:
\begin{eqnarray}
\label{decay-1}
\hat{R}_{s_f, s_\theta} &\equiv &
\langle N, s_f, {\bf P}_\theta-{\bf p}^\prime| \hat{t}
|\Theta^+, s_\theta, {\bf P}_\theta=0\rangle \, ,
\end{eqnarray}
and the photon density matrix elements $\rho_{s_\theta,s^\prime_\theta}$
in the $\Theta^+$ production can be obtained by squaring the amplitude
${\mathcal M}_{K} \left(\gamma N\rightarrow \Theta^+ K \right)$ of the
corresponding spin of the $\Theta^+$ and summing over the spin of the
nucleon and the helicity of the photon.

The transition operator $\hat{R}_{s_f, s_\theta}$ depends on the spin of
the particles involved.  Below is the list of the transition operator
$\hat{R}_{s_f, s_\theta}$ we use in this paper:
\begin{itemize}
\item $J^P = {\frac{1}{2}}^-$
\begin{center}
$\hat{R}_{s_f, s_\theta} = {\cal C} \delta_{s_f,s_\theta}$
\end{center}
\item $J^P = {\frac{1}{2}}^+$
\begin{eqnarray}
\hat{R}_{\frac{1}{2}, \frac{1}{2}} &=& \rm{cos} \, \theta \nonumber \\
\hat{R}_{\frac{1}{2}, -\frac{1}{2}} &=& -{\rm sin} \, \theta \,\,
e^{-i\phi} \nonumber \\
\hat{R}_{-\frac{1}{2}, \frac{1}{2}} &=& -{\rm sin} \, \theta \,\,
e^{i\phi} \nonumber \\
\hat{R}_{-\frac{1}{2}, -\frac{1}{2}} &=& -\rm{cos} \, \theta \nonumber
\end{eqnarray}

\item $J^P = {\frac{3}{2}}^+$
\begin{eqnarray}
\hat{R}_{\frac{1}{2}, \frac{3}{2}} &=& \frac{\rm sin \, \theta}{\sqrt 2}
\,\, e^{i\phi} \nonumber \\
\hat{R}_{\frac{1}{2}, \frac{1}{2}} &=& \sqrt{\frac{2}{3}}\rm{cos} \,
\theta \nonumber \\
\hat{R}_{\frac{1}{2}, -\frac{1}{2}} &=& -\sqrt{\frac{1}{6}}{\rm sin} \,
\theta \,\, e^{-i\phi} \nonumber \\
\hat{R}_{-\frac{1}{2}, \frac{1}{2}} &=& \sqrt{\frac{1}{6}}{\rm sin} \, \theta \,\,
e^{i\phi} \nonumber \\
\hat{R}_{-\frac{1}{2}, -\frac{1}{2}} &=& \sqrt{\frac{2}{3}}\rm{cos} \,
\theta \nonumber \\
\hat{R}_{-\frac{1}{2}, -\frac{3}{2}} &=& -\frac{\rm sin \, \theta}{\sqrt 2}
\,\, e^{i\phi} \nonumber
\end{eqnarray}

\item $J^P = {\frac{3}{2}}^-$
\begin{eqnarray}
\hat{R}_{\frac{1}{2}, \frac{3}{2}} &=& -\sqrt{\frac{3}{40}} {\rm sin} \, 2\theta
\,\, e^{i\phi} \nonumber \\
\hat{R}_{\frac{1}{2}, \frac{1}{2}} &=& -\sqrt{\frac{1}{10}}(3\,{\rm cos}^2 \,
\theta - 1) \nonumber \\
\hat{R}_{\frac{1}{2}, -\frac{1}{2}} &=& \sqrt{\frac{9}{40}} {\rm sin} \, 2\theta
\,\, e^{-i\phi} \nonumber \\
\hat{R}_{\frac{1}{2}, -\frac{3}{2}} &=& -\sqrt{\frac{3}{10}} {\rm sin}^2 \,\theta
\,\, e^{-2i\phi} \nonumber \\
\hat{R}_{-\frac{1}{2}, \frac{3}{2}} &=& \sqrt{\frac{3}{10}} {\rm sin}^2 \,\theta
\,\, e^{2i\phi} \nonumber \\
\hat{R}_{-\frac{1}{2}, \frac{1}{2}} &=& \sqrt{\frac{9}{40}} {\rm sin} \, 2\theta
\,\, e^{i\phi} \nonumber \\
\hat{R}_{-\frac{1}{2}, -\frac{1}{2}} &=& \sqrt{\frac{1}{10}}(3\,{\rm cos}^2 \,
\theta - 1) \nonumber \\
\hat{R}_{-\frac{1}{2}, -\frac{3}{2}} &=& -\sqrt{\frac{3}{40}} {\rm sin} \, 2\theta
\,\, e^{-i\phi} \nonumber
\end{eqnarray}

\end{itemize}

\end{document}